\documentclass[]{spieman}
\usepackage[]{graphicx}
\usepackage[]{deluxetable}
\usepackage{multicol}
\usepackage{rotating}
\usepackage{wrapfig,lipsum}
\usepackage{color}
\usepackage{subfig}

\title{Multi-Conjugate Adaptive Optics Simulator for the Thirty Meter Telescope: Design, Implementation, and Results}

\author[a]{Etsuko Mieda}
\author[b]{Jean-Pierre Veran}
\author[c]{Matthias Rosensteiner}
\author[d]{Paolo Turri}
\author[b]{David Andersen}
\author[b]{Glen Herriot}
\author[b]{Olivier Lardiere}
\author[e]{Paolo Spano}

\affil[a]{Subaru Telescope, 650 N. Aohoku Pl. Hilo, HI 96720, USA}
\affil[b]{National Research Council Herzberg Astronomy and Astrophysics, 5071 W Saanich Rd., Victoria, BC V9E 2E7, Canada}
\affil[c]{Max Planck Institute for Extraterrestrial Physics, Gie{\ss}enbachstra{\ss}e 1, 85741 Garching, Germany}
\affil[d]{University of California, Berkeley, Campbell Hall, \#501, Berkeley, CA 94720-3411, USA }
\affil[e]{Officina Stellare, Via Della Tecnica, 87/89, I-36030 Sarcedo (VI) - ITALY}

\begin{document}
\maketitle

%================
%=== ABSTRACT ===
%================
\begin{abstract}
We present a multi-conjugate adaptive optics (MCAO) system simulator bench, HeNOS (Herzberg NFIRAOS Optical Simulator). HeNOS is developed to validate the performance of the MCAO system for the Thirty Meter Telescope, as well as to demonstrate techniques critical for future AO developments. In this paper, we focus on describing the derivations of parameters that scale the 30-m telescope AO system down to a bench experiment and explain how these parameters are practically implemented on an optical bench. While referring other papers for details of AO technique developments using HeNOS, we introduce the functionality of HeNOS, in particular, three different single-conjugate AO modes that HeNOS currently offers: a laser guide star AO with a Shack-Hartmann wavefront sensor, a natural guide star AO with a pyramid wavefront sensor, and a laser guide star AO with a sodium spot elongation on the Shack-Hartmann corrected by a truth wavefront sensing on a natural guide star. Laser tomography AO and ultimate MCAO are being prepared to be implemented in the near future.
\end{abstract}


%====================
%=== INTRODUCTION ===
%====================
\section{INTRODUCTION}
\label{sec:intro}
NFIRAOS\cite{herriot:14} (Narrow Field InfraRed Adaptive Optics System) will be the first adaptive optics (AO) system to be deployed on the Thirty Meter Telescope (TMT). NFIRAOS is a multi-conjugate(MC) AO system equipped with six laser guide stars (LGSs), six LGS wavefront sensors (WFSs) made with Shack-Hartmann (SH) WFSs, two deformable mirrors (DMs), and a natural guide star (NGS) truth wavefront sensor (TWFS). NFIRAOS will provide near diffraction limited correction over 10 to 30 arcseconds and partial correction over 2 arcminutes in near infrared. The unprecedented scale of NFIRAOS prompts us to build the HeNOS (Herzberg NFIRAOS Optical Simulator) bench at National Research Council Herzberg for Astronomy and Astrophysics in Canada. The main purpose of this bench is to demonstrate the robustness and performance stability of a NFIRAOS-like MCAO system in realistic and varying conditions. 

In this paper, we present the design and implementation of HeNOS, and describe the bench capability to introduce AO development applications. Our work is organized as follows. A derivation of the bench parameters drawn from the NFIRAOS parameters and constraints is shown in \S \ref{sec:bench}. The components of the bench and how they are implemented on an optical bench are described in \S \ref{sec:benchcomp}. The new implementation of a TWFS made with a pyramid wavefront sensor (PWFS) can be found in \S \ref{sec:elon} and \ref{sec:pwfs}. \S \ref{sec:loop} describes three different AO modes that HeNOS offers today. Finally, we summarize the current bench status and discuss the future HeNOS plan in \S \ref{sec:sum}.

%=========================
%=== Bench Parameters ===
%=========================
\section{Bench Parameters}
\label{sec:bench}
The challenge in designing the HeNOS bench is to scale the AO system from a 30-meter telescope down to a bench size experiment. In this section, we describe how the bench parameters are derived from the NFIRAOS parameters. The parameter subscripts “N” and “H” refer to NFIRAOS and HeNOS, respectively. The earlier bench design development can be found in Refs.~\citenum{veran:12}.

The first set of constraints arises from the need to keep the cost down. The first cost-related constraint is the necessity to work at visible wavelength in order to simplify optics alignment and to use inexpensive CCD or CMOS detectors, such as our Point Grey Grasshopper CCDs, for WFS and imaging. %Note that the derivation discussed in the remaining of this section is done at $\lambda_{\rm H}$ = 0.690 $\mu$m; however, 
We have set $\lambda_{\rm H}$ = 0.670 $\mu$m as both our sensing and imaging (science) wavelength for the HeNOS bench to match the laser diode we had available as light source.
%after all. %As discussed later, the bench will end up operating at 0.670 $\mu$m

The second cost-related constraint is the necessity to work with relatively low order (thus economical) DMs. On HeNOS, we have two Alpao magnetic DMs. DM0 is a 11$\times$11 DM with $\sim$9 actuator pitches across the clear aperture ($n_{\rm H}$), conjugated to ground, and DM1 is 19$\times$19 DM with $n_{\rm H}\sim$16, which is conjugated to a high altitude. Both DMs have the same physical actuator pitch of 1.5mm. If we simulate a $D_{\rm H}$ = 30 m telescope with $n_{\rm H}=9$ across all NGS beams, diffraction-limited imaging could only be achieved with unrealistically weak turbulence. Diffraction-limited imaging is possible as long as the turbulence, $r_0$, is not significantly less than $d_{\rm H} = D_{\rm H}/n_{\rm H}$ at $\lambda_{\rm H}$. We have found that $D_{\rm H}$  = 8 m is about the largest feasible diameter that we can scale down to a bench size with our DMs. Consequently, we have set the telescope diameter to simulate with HeNOS to $D_{\rm H}$  = 8 m. It follows that the DM actuator pitch projected on this diameter is: $d_{\rm H} = D_{\rm H}/9 = 0.89$ m. 

The third cost-related constraint is the necessity to limit the field of view (FOV) of the system. Considering the size of the telescope we are trying to simulate (8 m) and the size of our DMs (13.5 mm footprint for an NGS beam), plus, in any case, the desire to work with off-the-shelf optical elements in an affordable size (i.e. 1-2 inches), we have found, after several iteration of the optical design, that the FOV had to be limited to FOV$_{\rm H}$ = 10.9 arcsec on the sky. This is significantly smaller than the NFIRAOS FOV, which is FOV$_{\rm N}$ = 120 arcsec on the sky. Note that this constraint was not integrated in the derivations from Refs.~\citenum{veran:12}.

Within the above set of constraints, we now set three objectives to guide our bench design:
\begin{itemize}
    \item Objective 1: HeNOS should have the ability to achieve diffraction limited imaging on axis. This constrains the simulated turbulence to have $r_{\rm 0,H}(\lambda_H) \sim d_{\rm H}$=0.89 m
    \item Objective 2: After MCAO correction, HeNOS and NFIRAOS should have the same PSF uniformity across their respective FOV. This is achieved if the ratio $\theta_2$/FOV is the same for HeNOS as for NFIRAOS, where $\theta_2$ is the generalized anisoplanatism angle after two DM corrections. This objective guarantees that the corrected field of HeNOS and NFIRAOS will look alike.
    \item Objective 3: HeNOS and NFIRAOS should have the same $\theta_0$/FOV ratio. This makes it just as hard for HeNOS and NFIRAOS to achieve Objective 2.
\end{itemize}
These objectives ensure that HeNOS has similar wide-field performance as NFIRAOS under similarly difficult turbulence conditions. These similarities ensure that the model used to predict the NFIRAOS performance will work under similar conditions as when it is used in the HeNOS configuration that is supposed to demonstrate its validity as discussed in the introduction. The similarities also ensure that the algorithms under tests (such as truth sensing, LGS-NGS tomography and PSF reconstruction) will be validated conditions similar to that in which they are expected to be used in NFIRAOS. 

\begin{deluxetable}{cccc}
\tabletypesize{\footnotesize}
\tablecolumns{4}
\tablewidth{0pc}
\tablecaption{NFIRAOS Parameters}
\setlength{\tabcolsep}{0.02in}
\tablehead{
\colhead{Abbrv} & 
\colhead{Name} &
\colhead{Unit} & 
\colhead{NFIRAOS}
}
\startdata
$D_{\rm N}$ & Telescope Diameter & [m] & 30 \\
$d_{\rm N}$ & DM actuator pitch & [m] & 0.5 \\
$\lambda_{\rm N}$ & Imaging wavelength & [$\mu$m] & 1.6 \\
$h_{\rm DM, N}$ & DM altitudes & [km] & [0, 11.2] \\
$r_0(0.5\mu m)$ & Fried parameter & [m] & 0.186 \\
$h_{\rm N}$ & Turbulence layer altitudes & [km] & [0, 0.5, 1, 2, 4, 8, 16] \\
$w_{\rm N}$ & Turbulence layer weights & & [0.4557, 0.1295, 0.0442, 0.0506, 0.1167, 0.0926, 0.1107] \\
$r_{0, \rm N}(\lambda_{\rm N})$ & Fried parameter at observing wavelength & [m] & 0.75 \\
$\theta_{0, \rm N}(\lambda_{\rm N})$ & Anisoplantic angle & [arcsec] & 9.4 \\
$\theta_{2, \rm N}(\lambda_{\rm N})$ & Generalized anisoplantic angle after 2 DM correction & [arcsec] & 34.6 \\
\enddata \label{tab:para}
\end{deluxetable}
In order to achieve our three objectives, we start with $r_{\rm 0,H}(\lambda_H)$ = 0.89 m. The isoplanatic angle is obtained as $\theta_0$ = 0.3147$r_0/h_0$ where $h_0$ is the weighted average altitude of the turbulence. For the NFIRAOS median profile given in Table \ref{tab:para}, $h_{\rm 0, N}\sim$ 5.2 km. Objective 3 can be written as:
\begin{equation}
    \frac{\theta_{\rm 0, H}(\lambda_{\rm H})}{\theta_{\rm 0, N}(\lambda_{\rm H})}=\frac{r_{\rm 0, H}(\lambda_{\rm H})}{r_{\rm 0, N}(\lambda_{\rm H})}\frac{h_{\rm, 0, N}}{h_{\rm 0, H}}=\frac{FOV_{\rm H}}{FOV_{\rm N}}
\end{equation}
With FOV$_{\rm H}$ constrained to 10.9 arcsec, the only parameters that can be adjusted is the mean turbulence altitude on the HeNOS bench ($h_{\rm 0,H}$). This leads to $h_{\rm 0,H} = 12.9 h_{\rm 0,N}$ = 67.0 km.

Objective 3 can be achieved by keeping the same turbulence profile as NFIRAOS, but stretching it by a factor $f_s$ = 12.9. Then Objective 2 can be satisfied by simply increasing the distance between DM0 and DM1 by the same factor. However, because the diameter of the meta-pupil on DM1 (ensemble of the footprint of all NGSs within the FOV -- each NGS has a 13.5mm footprint) is larger than DM1 clear aperture, which is 24.5 mm, $f_s$=12.9 does not work well. The maximum meta-pupil diameter at the conjugate altitude of DM1 is:
\begin{equation}
    D_{\rm meta} = D_{\rm H} \frac{24.5}{13.5} = 14.5 \textrm{ [m]}.
\end{equation}
It follows that the maximum conjugate altitude of DM1 is:
\begin{equation}
    h_{\rm DM1, H} = \frac{D_{\rm meta}-D_{\rm H}}{\textrm{FOV}_{\rm H}(\lambda_{\rm H})}=123 \textrm{ [km]},
\end{equation}
which corresponds to a maximum stretch factor of
\begin{equation}
    f_s = h_{\rm DM1, H}/h_{\rm DM1, N}=11.
\end{equation}

The only way to satisfy Objective 2 and 3 with the maximum stretch factor derived above is to reduce $r_{\rm 0,H}$ to:
\begin{equation}
    r_{0, \rm H}(\lambda_{\rm H}) = 0.751 \textrm{ [m]}.
\end{equation}

This $r_{\rm 0, H}$ value is still close to $d_{\rm H}$, and thus Objective 1 is still achieved. It is worth noting that 
$r_{\rm 0, H}$ actually does not depend on $\lambda_{\rm H}$.%; the stretch factor however is inversely proportional to $\lambda_{\rm H}$. 
Also interesting is that the maximum stretch factor is very close to the ratio between the NFIRAOS and the HeNOS FOVs. This coincidence was not planned. It arises from the difference in size of the two DMs. In retrospect, we realize that if the second DM had been significantly smaller, we might not have been able to scale HeNOS properly.

The fitting error  is given by%does not depend on $\lambda_{\rm H}$:
\begin{equation}
\sigma_{\rm fit, H} = \frac{\lambda_{\rm H}}{2\pi}\sqrt{0.23\left(\frac{d_{\rm H}}{r_{0, \rm H}(\lambda_{\rm H})}\right)^{5/3}}= 59%60.0
\textrm{ [nm] rms},%.
\end{equation}
where $\lambda_{\rm H}=670$ nm. The same formula gives $\sigma_{\rm fit, N}$ = 87.7 nm rms for NFIRAOS. The agreement between the bench and NFIRAOS could be improved by reducing the NFIRAOS reference wavelength. %On the other hand, the impact on the Strehl ratio is wavelength dependent:
The Strehl ratio is given by:
\begin{equation}
    SR_{\rm fit, H} = \exp{\left(-0.23\left(\frac{d_{\rm H}}{r_{0, \rm H}(\lambda_{\rm H})}\right)^{5/3}\right)}=0.74.
\end{equation}
The same formula gives $SR_{\rm fit, N}$ = 0.89 for NFIRAOS.

\begin{deluxetable}{cccc}
\tabletypesize{\footnotesize}
\tablecolumns{4}
\tablewidth{0pc}
\tablecaption{HeNOS Parameters}
\setlength{\tabcolsep}{0.02in}
\tablehead{
\colhead{Abbrv} & 
\colhead{Name} &
\colhead{Unit} & 
\colhead{HeNOS}
}
\startdata
$D_{\rm H}$ & Telescope Diameter & [m] & 8 \\
$d_{\rm H}$ & DM actuator pitch & [m] & 0.89 \\
$\lambda_{\rm H}$ & Imaging wavelength & [$\mu$m] & 0.67 \\
$h_{\rm DM, H}$ & DM altitudes & [km] & [0, 123] \\
$f_s$ & Scaling factor & & 11 \\
%$r_0(0.5\mu m)$ & Fried parameter & [m] & 0.186 \\
%$h_{\rm H}$ & Turbulence layer altitudes & [km] & [0, 0.5, 1, 2, 4, 8, 16] \\
%$w_{\rm H}$ & Turbulence layer weights & & [0.4557, 0.1295, 0.0442, 0.0506, 0.1167, 0.0926, 0.1107] \\
$r_{0, \rm H}(\lambda_{\rm H})$ & Fried parameter at observing wavelength after applying $f_s$ & [m] & 0.751 \\
$\theta_{0, \rm H}(\lambda_{\rm H})$ & Anisoplantic angle & [arcsec] & 0.854\\
%$\theta_{2, \rm H}(\lambda_{\rm H})$ & Anisoplantic angle after 2 DM correction & [arcsec] & 34.6 \\
\enddata \label{tab:hpara}
\end{deluxetable}
The bench parameters described above are summarized in Table \ref{tab:hpara}. Note that in Ref.~\citenum{veran:12}, the derived parameters were different, especially the stretch factor, which was only 4.2. This is because the need to limit the FOV to 10.9 arcsec was not recognized in Ref.~\citenum{veran:12}.

In Ref.~\citenum{veran:12}, we simulated NFIRAOS and HeNOS with 4 LGSs, and Table 5 of Ref.~\citenum{veran:12} summarizes the results. The total RMS wavefront error for HeNOS and NFIRAOS are 93 nm and 156 nm, leading to a delivered Strehl ratio of 0.49 and 0.69, respectively, at their respective wavelength. These simulation results remain valid even if the stretch factor is different because the FOV and asterism diameter have been scaled accordingly. While HeNOS has lower wavefront error, the lower wavelength increases the sensitivity of the Strehl ratio, and therefore departure from nominal performance should be easily detectable.

%Now that the turbulence profile and DM architecture have been defined for HeNOS, we can turn to the LGS constellation. The asterism is set to a 4.5 arcsec square on the sky. In order to preserve the LGS cone angle through the turbulence (and therefore keep a realistic focal anisoplanatism), we also multiply the nominal range of the LGS by the stretch factor (90 km x 11 = 990 km). This value provides a reasonable overlap even at the highest turbulence layer, now at 16 km x 11 = 176 km. The LGS footprint at this layer is 8 m x (990 – 176)/990 = 6.6 m, and the footprint of two LGSs, 4.5 arcsec apart, are separated by 3.8 m, leaving 1.8 m or about 30\% overlap.

%========================================
%=== Bench Components and Calibration ===
%========================================
\section{Bench Components and Calibration}
\label{sec:benchcomp}
\begin{figure}[t]
\centering
\includegraphics[width=0.9\textwidth]{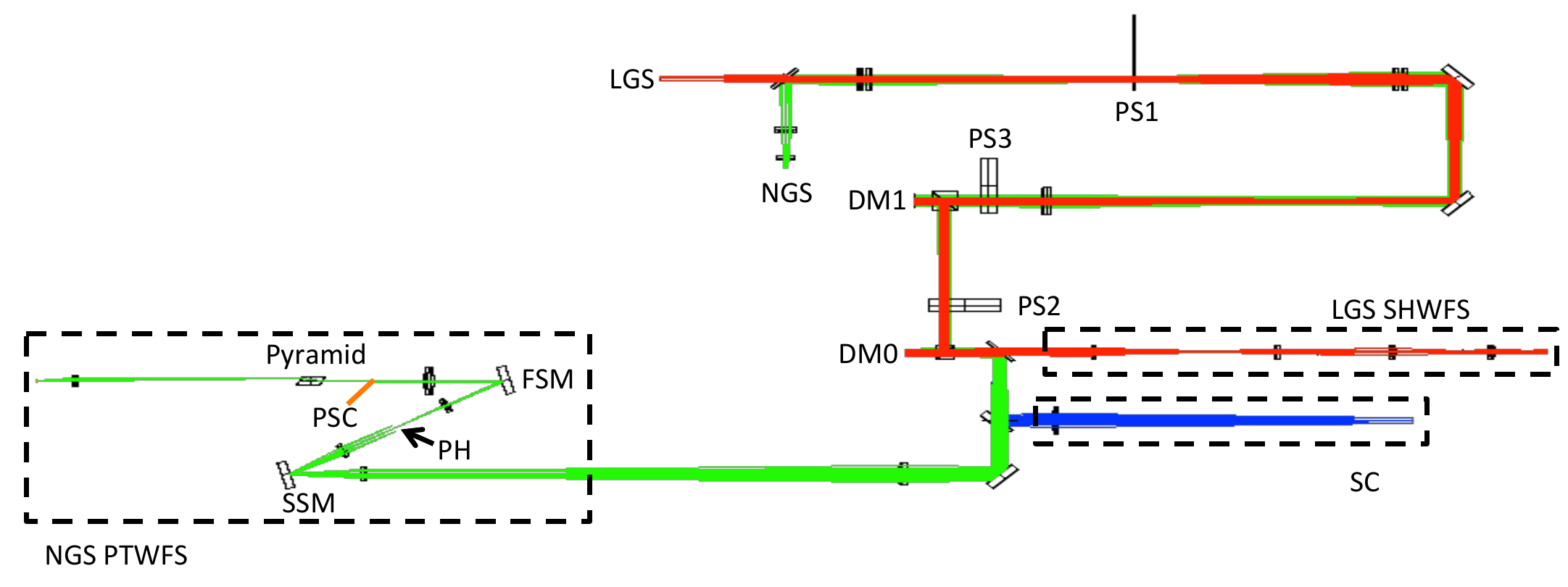}
\caption{The optical paths of the HeNOS bench. Abbreviations are described in Table \ref{tab:benchpara}. The bench consists of four LGSs in a 2 by 2 configuration, a grid of NGSs, two DMs conjugated to 0 and 12 km, three PSs conjugated to 0.6, 5.2, and 16.3 km, one SHWFS (red) simultaneously measuring four LGSs, one SC (blue), and one TWFS made with a double pyramid (green). To calibrate the PWFS performance, one more science camera focused on NGS, called PSC (orange), is also added.}
\label{fig:henos}
\end{figure}
The bench is built based on the bench parameters described in \S \ref{sec:bench} and Table \ref{tab:hpara}. Figure \ref{fig:henos} shows its most updated optical paths, and Table \ref{tab:benchpara} summarizes the abbreviated components. In order to understand the experimental results on the HeNOS bench and use them for NFIRAOS and other AO developments, it is important to know the precise dimensions of the bench. The actual as-built parameters may deviate from the design parameters, but only to a small degree.  In this section, we describe each component of the HeNOS bench in detail and report the calibration results. The earlier bench development and calibration results can be found in Refs.~\citenum{turri:14} and \citenum{rosensteiner:15}. NFIRAOS will be operated at -30 $^\circ$C, and its components will be tested under the low temperature; however, the HeNOS bench is designed to work at only room temperature, and we do not talk about cold environment here.
\begin{deluxetable}{ccl}
\tabletypesize{\footnotesize}
\tablecolumns{3}
\tablewidth{0pc}
\tablecaption{Bench Components}
\setlength{\tabcolsep}{0.02in}
\tablehead{
\colhead{Abbrv} & 
\colhead{Name}&
\colhead{Description}
}
\startdata
LGS & Laser Guide Star & 2 by 2 configuration lasers whose separation defines 4.5 arcsec on the sky. \\
NGS & Natural Guide Star & Creates a grid of NGSs, previously by microlens array, future by pinhole mask.\\
DM0 & Deformable Mirror 0 & ALPAO DM with 97 actuators at ground (0 km).\\
DM1 & Deformable Mirror 1 & ALPAO DM with 277 actuators at high altitude (12 km).\\
PS1 & Phase Screen 1 & UCSC (paint spraying) PS at ground layer (0.6 km).\\
PS2 & Phase Screen 2 & Lexitex (index matching) PS at middle layer (5.2 km).\\
PS3 & Phase Screen 3 & Lexitex (index matching) PS at high layer (16.3 km).\\
FSM & Fast Steering Mirror & Newport FSM-300.\\
PH & Pinhole & 500 $\mu$m pinhole at NGS focus to block all but one NGS.\\
SSM & Star Selection Mirror & Zaber motorized gimbal mount and mirror.\\
SHWFS & Shack-Hartmann WFS & 30 by 30 subapeture SHWFS with Pointgrey Grasshopper (2448x2048, 3.45$\mu$m pixel).\\
SC & Science Camera & Andor sCMOS Zyla (2048x2048, 6.5$\mu$m pixel) currently at LGS focus. \\
PTWFS & Pyramid Truth WFS & 76 pixel diameter pupil PWFS with Pointgrey Flea (638 x 488, 5.6 $\mu$m pixel).\\
PSC & Pyramid Science Camera & Pointgrey Grasshopper (2448 x 2048, 3.45 $\mu$m pixel) at NGS focus.
\enddata \label{tab:benchpara}
\end{deluxetable}

%===========
%=== LGS ===
%===========
\subsection{Laser Guide Star}
To measure the bench's as-built parameters, we fix one parameter and derive all others. We chose the LGS asterism size as the fixed parameter since the LGSs are mounted in solid holes on a metallic plate, which are sturdy and least likely to change over time. 

NFIRAOS will project its six LGSs within a radius of 35 arcsec on sky (one at the center and five on the circle), whose stretched correspondence on HeNOS is 6.4 arcsec. For simplicity, we designed the HeNOS bench with four fixed 
\begin{wrapfigure}{r}{0.5\textwidth}
\includegraphics[width=0.5\textwidth]{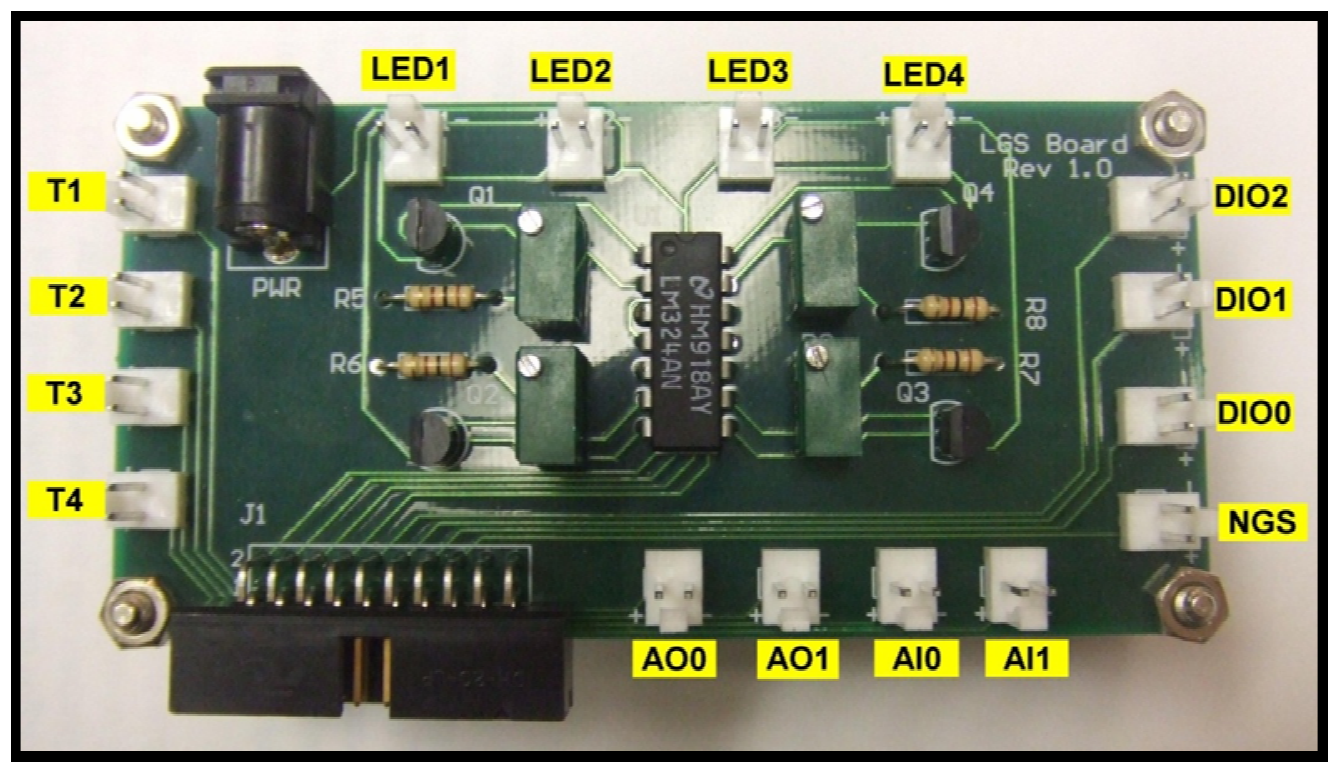}
\caption{A photo of the LGS Board. It provides an interface for the four LGSs, NGS, and SHWFS camera trigger.}
\label{fig:lgsboard}
\end{wrapfigure}
LGSs with a square asterism. For a square asterism, the corresponding side length is 4.5 arcsec. In order to preserve the LGS cone angle through the turbulence (and therefore keep a realistic focal anisoplanatism), we also multiply the nominal range of the LGS by the stretch factor (90 km x 11 = 990 km). This value provides a reasonable overlap even at the highest turbulence layer, now at 16 km x 11 = 176 km. The LGS footprint at this layer is 8 m x (990 - 176)/990 = 6.6 m, and the footprint of two LGSs, 4.5 arcsec apart, are separated by 3.8 m, leaving 1.8 m or about 30\% overlap.

All four LGSs are made with single-mode fiber laser diodes from Thorlabs (LPS-675-FC), whose optical output power is 2.5 mW. The laser diodes are controlled by a combination of an ADLINK data acquisition card (DAQe-2502) and a custom printed circuit board manufactured by Alberta Printed Circuit (Figure \ref{fig:lgsboard}). Using the DAQ card, the LGS timing and intensity can be controlled using computer commands.

%===========
%=== NGS ===
%===========
\subsection{Natural Guide Star}
\label{sec:ngs}
\begin{figure}[b]
\centering
%\begin{wrapfigure}{r}{0.6\textwidth}
\includegraphics[width=0.6\textwidth]{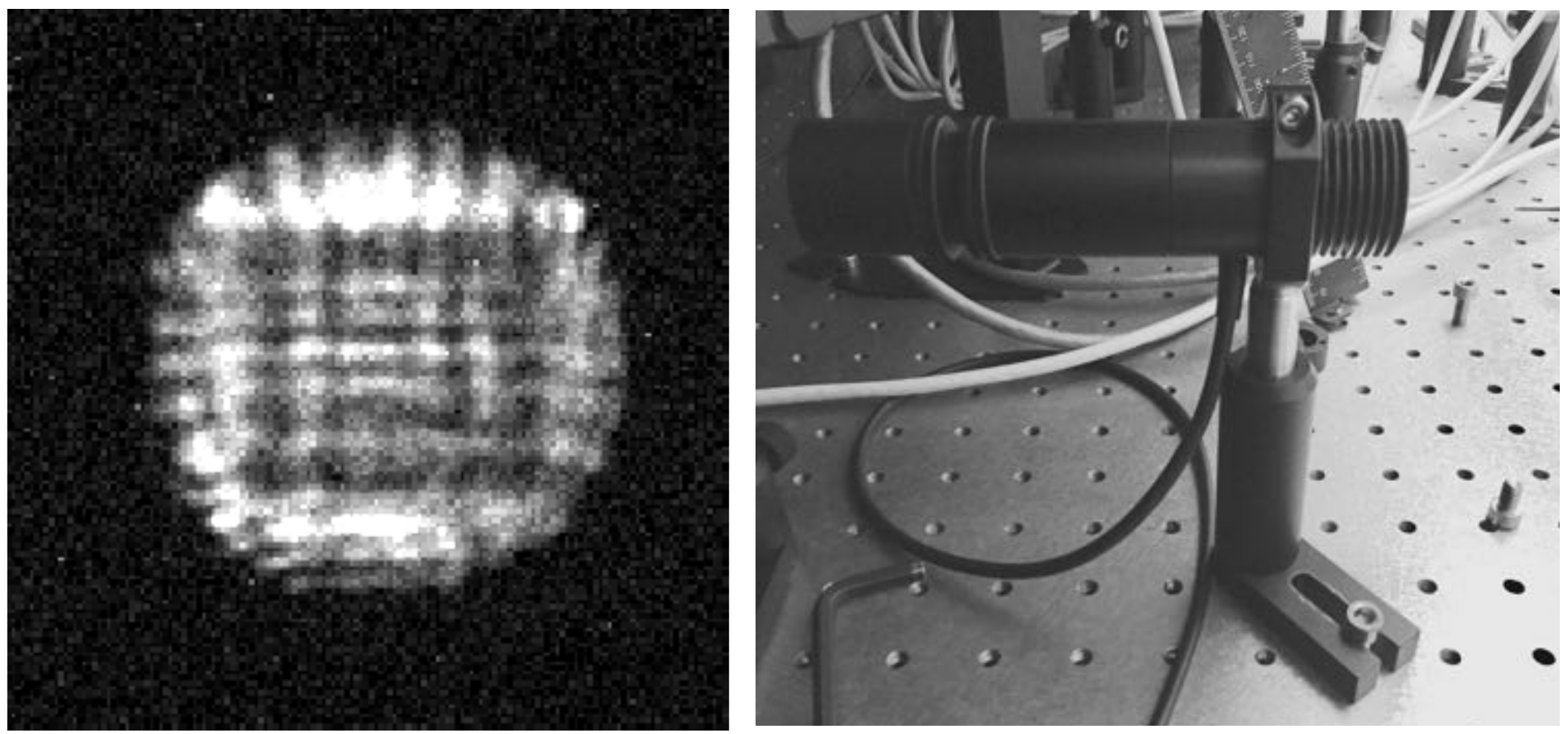}
\caption{Left: image of a grid pattern pupil from the old NGS light source with a MA. The grid pattern is likely created by the concentrated light from the gaps between microlenses on top of the pupil. Right: New NGS light source test setup on the bench. The new design uses a pinhole mask instead of a MA.}
\label{fig:pup}
%\end{wrapfigure}
\end{figure}
To evaluate the AO performance across the entire science field, many PSFs on the science camera are needed. We originally created a grid of NGSs using a laser diode (same one as LGSs) and a microlens array (MA). The use of a MA was an easy way to produce many PSFs at the focus, but it creates a grid intensity pattern on the pupil (Figure \ref{fig:pup} left panel). This grid pattern is likely created by the concentrated light from the gaps between microlenses on top of the pupil.

We are currently working on a new NGS design without a MA. The new design consists of a powerful LED, a collimating lens, a diffuser, and a pinhole mask in a lens tube. The right panel of Figure \ref{fig:pup} shows the test setup. To produce a PSF, the size of the pinhole has to be smaller than the diffraction limit of the bench, which is $\sim$ 10 $\mu$m at the NGS position in Figure \ref{fig:henos}. The HeNOS bench uses many beamsplitters, including two beamsplitter cubes in front of the DMs, which waste a large fraction of photons, and the tiny pinhole blocks even more photons unlike the MA. We are adjusting the position of the collimating lens and pinhole mask to maximize photon count at the pinhole mask, and adjusting the beamsplitters using different reflection-transmission ratios to give more weight to the NGS path.

For a temporary solution, we are using the laser diode itself, without MA nor pinhole mask, as a single NGS. The size of the laser diode is small enough to create a diffraction-limited PSF. Because it is only one PSF, we cannot evaluate AO performance over a whole FOV on the science camera, but it is used for a PTWFS experiment (see \S \ref{sec:pwfs} and \ref{sec:loop} for more about PTWFS). For performance evaluation, we place the science camera at the LGS focus and use LGSs as PSFs (see \S \ref{sec:scicam}).

We are also looking into a design where the pinhole mask can be easily replaced by a single pinhole of a large radius. Using a bigger pinhole, we can experiment a wavefront sensing on an extended object using PWFS. This experiment is not a direct simulation of NFIRAOS; however, it is useful to advance observations. See Refs.~\citenum{mieda:17} for PWFS simulations with extended guide objects and their science applications. Using a larger pinhole, the NGS is no longer diffraction-limited, and thus we cannot evaluate the performance directly from PSF. But since we have a separate LGS-SHWFS, we can check its performance using SHWFS measurements.

%=========================
%=== Deformable Mirror ===
%=========================
\subsection{Deformable Mirror}
\label{sec:dm}
We use two magnetic DMs from ALPAO, whose specifications are listed in Table \ref{tab:dm}. Since both DMs have the same actuator pitch, they are located in the same collimated space, with DM1 conjugated to the high altitude and DM0 to the ground (see Figure \ref{fig:henos} for their locations).  Beamsplitter cubes are placed in front of both DMs so that the incident beam hits the DM surfaces normally. These cubes simplify the bench design but waste a significant fraction of photons (see \S \ref{sec:ngs} for their drawback).

We determine the actuator spacing and the altitude conjugation of the mirrors by looking at the poke matrix taken with a SHWFS (see \S \ref{sec:shwfs} for SHWFS details). From the poke matrix, we measure the actuator positions in the WFS geometry and fit a uniform grid to them using a least square fit. The center of the fit is used to better align the SHWFS to avoid vignetting of the pupil. The actuator spacing derived from the scaling of the fit to DM0 is 0.914 m, and the conjugation altitude of DM1 estimated from the shear of the metapupils, knowing the LGS separation is 4.5 arcsec, is 121km with the stretch factor, $f_s$, applied (12 km without).
\begin{deluxetable}{ccc}
\tabletypesize{\footnotesize}
\tablecolumns{3}
\tablewidth{0pc}
\tablecaption{DM specifications}
\setlength{\tabcolsep}{0.02in}
\tablehead{
\colhead{Name} &
\colhead{DM0} & 
\colhead{DM1}
}
\startdata
Model & Hi-Speed DM97-15 & Hi-Speed DM227-15 \\
Actuator & 97 (11 $\times$ 11 square) & 277 (19 $\times$ 19) \\
Actuator pitch & 1.5 mm & 1.5 mm \\
Pupil diameter & 13.5 mm (9 pitches) & 24.5 mm (16.3 pitches) \\
Bandwidth & $>$ 750 Hz & $>$ 800 Hz \\
Settling time ($\pm$ 5\%) & 1.0 ms & 1.0 ms
\enddata \label{tab:dm}
\end{deluxetable}

%====================
%=== Phase Screen ===
%====================
\subsection{Phase Screen}
\label{sec:ps}
A key requirement of the HeNOS bench is the ability to generate realistic turbulence. To achieve this, we adopt well calibrated turbulence screens: two screens from Lexitek\cite{mantravadi:04} and one from UCSC\cite{rampy:10}. Lexitek screens use an index matching technique, where the turbulence profile can be flexibly designed with any $r_0$, but is expensive. The UCSC screen uses the acrylic paint spraying technique, where the cost is less, but only small $r_0$ is available. 

The two Lexitek phase screens (PS2 and PS3 in Figure \ref{fig:henos}) are placed in the same collimated space as the two DMs. PS3 just before DM1 has the bigger $r_0$ and simulates the highest turbulence layer, and PS2 in between the two DMs simulates the middle layer turbulence. Because the UCSC screen (PS1) is conjugated to the ground, which is occupied by DM0, a separate pupil position is created (see PS1 position in Figure \ref{fig:henos}). At this position, the designed beam size is 10 mm.

Two Lexitek PSs are mounted on Lexitek motorized rotary stages (LS-100), and the UCSC PS is controlled by a Galil motion controller. To measure wind speed, we measure the distance between the optical axis and the rotation center. The relationship between the physical dimensions and the simulated ones is given by the simulated telescope size and the physical pupil size in the collimated space (see \S \ref{sec:telsize} for the simulated telescope size). From this relationship, the circumferences of three PSs are 295 (PS1), 97 (PS2), and 92 m (PS3). Note that due to the rotational movement, the speed is never uniform across the metapupil. 

The altitudes of the PSs are calculated from the physical positions of the optical surfaces. Knowing the physical position (actual physical distance between DM0 and DM1 on the bench measured by a ruler) and the simulated altitude (\S \ref{sec:dm}) of DM1, the scaling factor relating the bench and the altitude of the atmosphere in the collimated space, where DM0, DM1, PS2, and PS3 are placed, is 28.22 mm/km. The scaling factor is proportional to the square of the aperture size ratio, and thus the PS1 altitude is $28.22 * (10/13.5)^2 = 15.8$ mm/km, where aparture sizes are 10 m at PS1 and 13.35 mm at the DM0's position.

For turbulence power measurements, we derive the Fried parameter using two independent methods: 1) the full-width-half-maximum (FWHM) of PSFs on the SC, assuming Kolomogorov statistics ($r_{\rm 0, SC}$ on Table \ref{tab:r0}), and 2) the standard deviation of the wavefront reconstructed from WFS slopes using a CuReD\cite{rosensteiner:12} reconstructor ($r_{\rm 0, WFS}$ on Table \ref{tab:r0}). For method 1, we use the NGS light source with a MA to create many stars for better statistics and take long exposure PSF images on the SC. For method 2, we take the slope measurements on the SHWFS. In both cases, we use one PS and take data on 100 different positions across the PS at a time. Data without PSs are also taken for reference. The resultant Fried parameters are listed in Table \ref{tab:r0} along with the nominal value used by the manufacturer ($r_0$). Note that the measurements of the high altitude PS need to be corrected for the cone effect. All PS measurements are well within 10 \% of the nominal value, but $r_{0, \rm WFS}$ is larger than $r_{0, \rm SC}$. This is because the WFS is blind to the highest spatial frequencies of the turbulence, and thus we believe $r_{0, \rm SC}$ is more accurate. The relative powers of the PSs in terms of $\sigma^2 \propto r_0^{-5/3}$ are 74.3 \%, 17.4 \%, and 8.2 \%.

\begin{deluxetable}{cccccc}
\tabletypesize{\footnotesize}
\tablecolumns{6}
\tablewidth{0pc}
\tablecaption{Fried Parameter}
\setlength{\tabcolsep}{0.02in}
\tablehead{
\colhead{Phase Screen} &
\colhead{$r_{0, \rm SC}$ (670 nm)} & 
\colhead{$r_{0, \rm SC}$ (500 nm)} & 
\colhead{$r_{0, \rm WFS}$ (670 nm)} &
\colhead{$r_{0, \rm WFS}$ (500 nm)} &
\colhead{$r_0$ (500 nm)} \\
 & [m] & [m] & [m] & [m] & [m]
}
\startdata
PS1 & 0.99 & 0.70 & 1.15 & 0.81 & 0.74 \\
PS2 & 2.37 & 1.67 & 2.83 & 1.99 & 1.61 \\
PS3 & 3.72 & 2.62 & 3.80 & 2.68 & 2.79 \\
All & 0.83 & 0.58 & \nodata & \nodata & \nodata
\enddata \label{tab:r0}
\end{deluxetable}

%=======================================
%=== Shack-Hartmann Wavefront Sensor ===
%=======================================
\subsection{Shack-Hartmann Wavefront Sensor}
\label{sec:shwfs}
Our WFS is a custom-made SHWFS with a square MA (300 $\mu$m pitch) and a Point Grey Grasshopper CCD (2448 $\times$ 2048 array with 3.45 $\mu$m $\times$ 3.45 $\mu$m pixel). The FOV of the lenslets is large enough to separately see all four LGSs, and thus a single detector is used to sense all four LGSs simultaneously. Depending on the experiment, individual LGS can be used as well if separate measurements are required. Our SHWFS has many subapertures (30 across the pupil) to sample the elongation finely at each distance away from the center.

We identify the spots created by the MA, and the average separation of the LGSs is 20.58 $\pm$ 0.51 pixel, which translates to the WFS pixel size being 0.22 arcsec. We fit a uniform grid to the SHWFS spots and define subapertures, including the ones outside of the illuminated zone. We then determine the pupil size by matching the measured illumination pattern and a modeled illumination pattern. The modeled illumination is the percentage coverage map of subapertures when a perfect circle is projected on a square subapertures. Applying this method to four LGSs separately, we find the radius of the pupil on the SHWFS to be 15.226 %$\pm$ 0.016 
subapertures.

Assuming a small aberration approximation, we estimate a WFS fitting error using Strehl ratio (SR) of NGS  First we measure SR$_1$ using flat mirrors instead of DMs and no phase screens. Then we measure SR$_2$ using DM0 and a ground layer PS after closing loop. The error is estimated as:
\begin{equation}
    \sigma^2_{\rm AO} = \ln{\frac{SR_1}{SR_2}}.
\end{equation}
Using median SRs, $\sigma_{\rm AO}$ = 33 nm. 

More thorough derivation of the error budget can be found in Ref.~\citenum{turri:17}.

%======================
%=== Science Camera ===
%======================
\subsection{Science Camera}
\label{sec:scicam}
We originally used a Pointgrey Grasshopper (2448$\times$2048, 3.45 $\mu$m pixel), the same model as the one used in the SHWFS, for the science camera; however, we realized that the Grasshopper has two readout channels which create background offset between the two sides of the detector. %This is a problem when we do the PSF reconstruction experiment. 
When a PSF falls near this divided region, it adds spatially different noise properties and complicates the modeling. We also noticed that the offset varies from time to time. We thus replaced the camera with an Andor sCMOS Zyla (2048$\times$2048, 6.5 $\mu$m pixel) which has a low noise and almost uniform background. The Andor sCMOS has a similar number of pixels but with larger pixels, and we realigned the lenses in front of it to have a same plate scale as when a Grasshopper was used.

Due to the NGS light source problem (\S \ref{sec:ngs}), currently we do not have a proper NGS light source on the bench. While we work on new solutions, for a temporary fix, we locate the science camera at the focus of the LGSs, and use them as science targets. At this location, their plate scale is 5.6 milli-arcsec.

%======================
%=== Telescope Size ===
%======================
\subsection{Telescope Size}
\label{sec:telsize}
%\begin{figure}[t]
%\centering
\begin{wrapfigure}{r}{0.6\textwidth}
\includegraphics[width=0.6\textwidth]{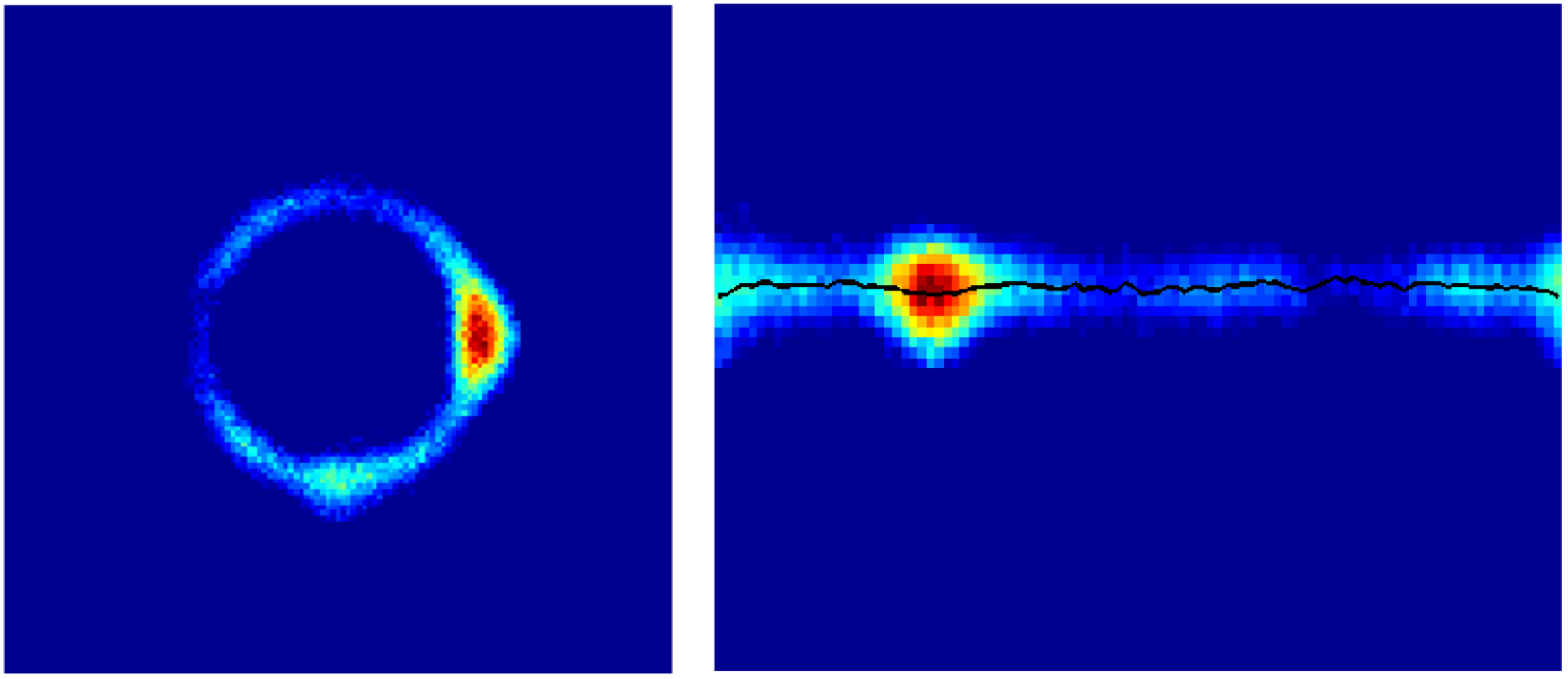}
\caption{To measure the simulated telescope size, many PSFs on the SC created by NGS source with a MA are stacked. The core-masked stacked PSF is shown on the left. The stacked PSF is then linearized by transforming from polar to Cartesian coordinate, and the radius of the ring is the median position of the central line, shown on the right.}
\label{fig:telsize}
\end{wrapfigure}
%\end{figure}
The simulated on-sky telescope diameter was measured in the earlier stages of bench development using the PSF Airy rings. The bench had flat mirrors instead of DMs, a NGS light source with a MA, no PSs, and an iris at PS1's position. We used a 3 mm iris to have the first ring distant enough from the core while keeping a sufficiently bright aperture, to measure its size. 

We took an image on the science camera and identified PSFs on the image. To obtain a high SNR, we stack all PSFs by aligning the brightest pixels (left on Figure \ref{fig:telsize}, the core masked out). To measure the diameter of the ring, we linearized the PSF image by transforming it from polar to Cartesian coordinates (right on Figure \ref{fig:telsize}). The radius of the ring is the median position of the central line and is 1.64 $\lambda/D$, which corresponds to an aperture of 2.44 m on sky. Scaling it to the 10 mm aperture at the first pupil's position (PS1's position), the HeNOS telescope size is 8.13 m.

%==============================
%=== SHWFS Spot Elongation and Truth Wavefront Sensor ===
%==============================
\subsection{SHWFS Spot Elongation and Truth Wavefront Sensor}
\label{sec:elon}
\begin{figure}[b]
\centering
\includegraphics[width=0.7\textwidth]{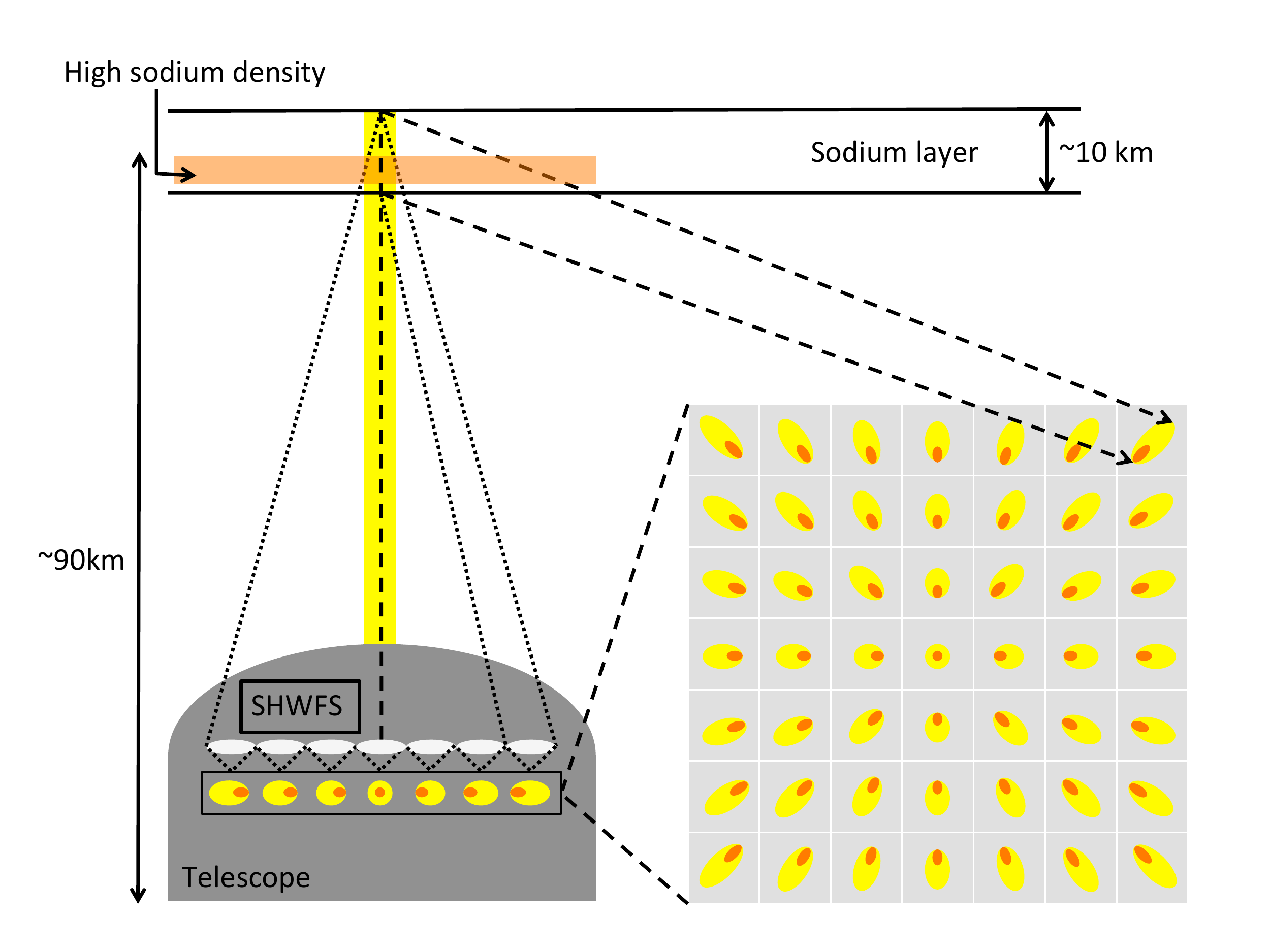}
\caption{Schematic of LGS SHWFS spot elongation. The example spots shown here are for a center-launched LGS. The subapertures further away from the center see the Na layer with an angle compared to those near the center, resulting in more radially elongated spots. This offset from the center increases with telescope size, and is particularly severe for ELTs. The layer shown as orange in the sodium layer represents high Na density, and its location in SHWFS spots are also shown as orange. Because of the high Na density, the orange part of the spots has higher flux return, which would be confused as a radial spot shift.}
\label{fig:twfs}
\end{figure}
\begin{figure}[t]
\centering
%\begin{wrapfigure}{r}{0.7\textwidth}
\includegraphics[width=0.7\textwidth]{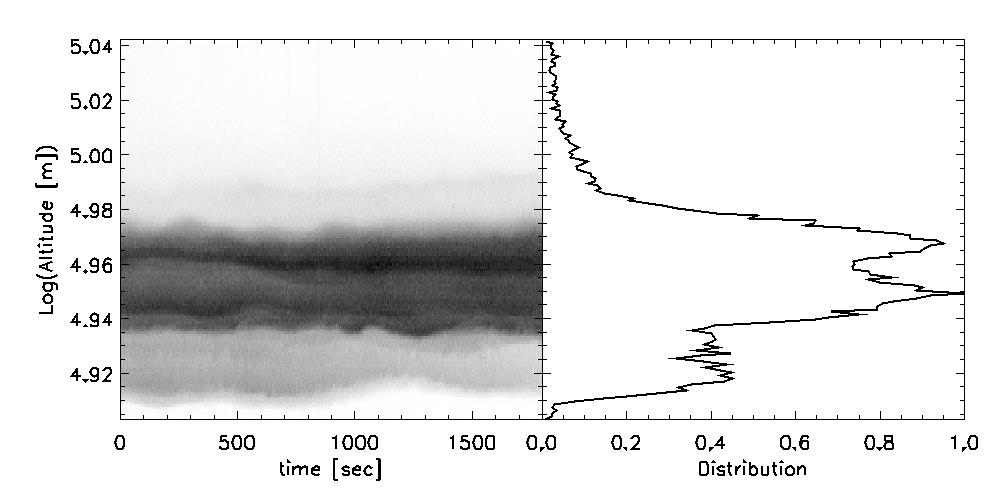}
\caption{Left: Empirically obtained Na profiles as functions of time over 1800 seconds. Darker color for higher density. Right: 1D plot of Na profile at $t = 0$ second.}
\label{fig:naprofile}
%\end{wrapfigure}
\end{figure}
When a Sodium (Na) laser is used as a LGS, the spots on a SHWFS are radially elongated due to the finite thickness of the Na layer. When the Na layer profile (i.e., height, thickness, and density distribution) changes, the photon 
\begin{wrapfigure}{r}{0.5\textwidth}
\includegraphics[width=0.5\textwidth]{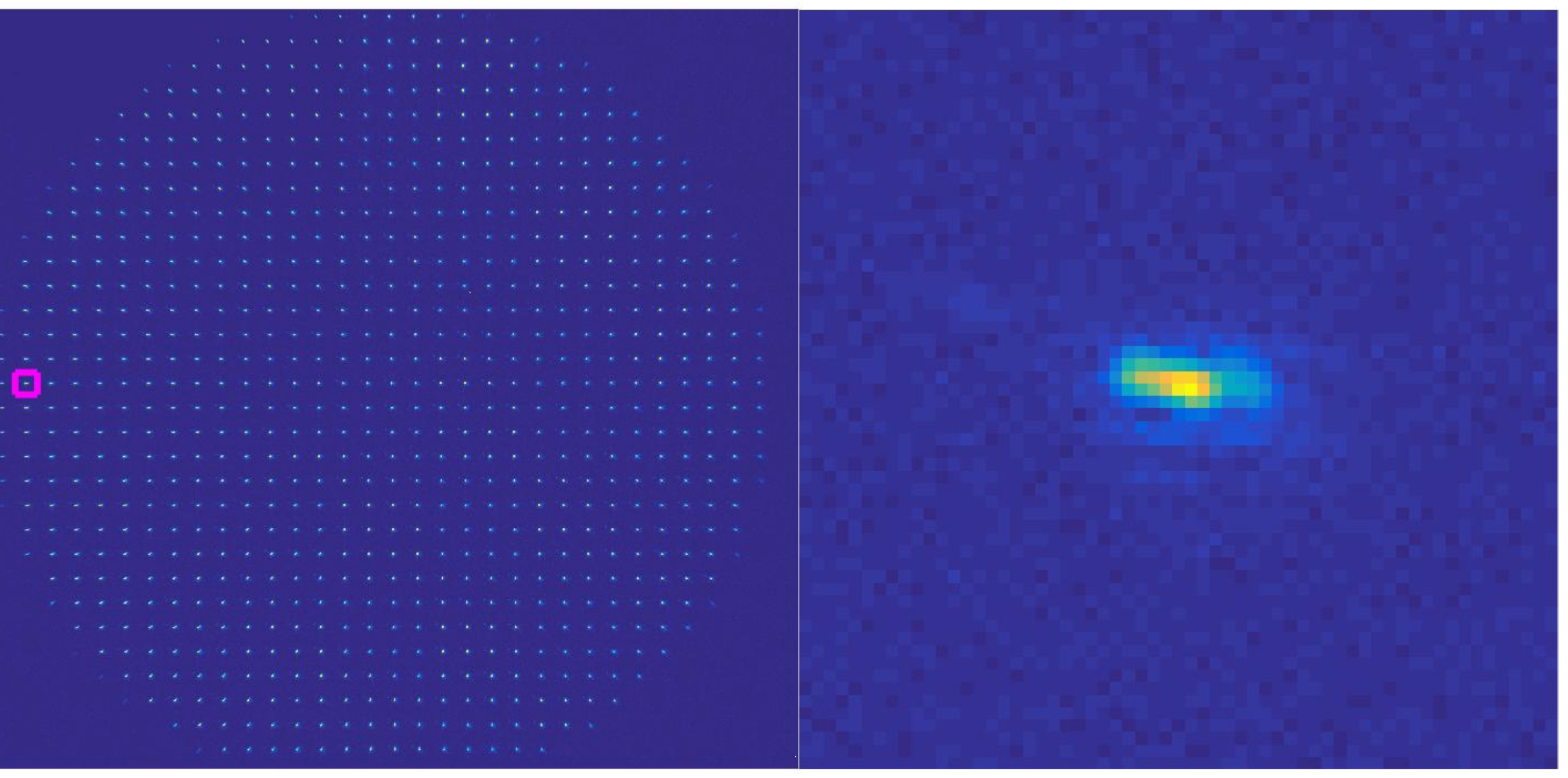}
\caption{Simulated elongated SHWFS spots when Na profile at $t=0$ (right on Figure \ref{fig:naprofile}) is applied. Full frame SHWFS camera image is shown on the left, and one zoomed spot in a magenta box is shown on the right.}
\label{fig:elon_wfs}
\end{wrapfigure}
distributions in the elongated spots also change. This introduces an additional centroid shift when centroding is applied \cite{vandam:06}. This spot elongation problem is more severe as the diameter of a telescope increases. Figure \ref{fig:twfs} explains how the thickness of the Na layer produces more elongated spots with larger telescopes, and why the fluctuation in Na profile causes aberrations. See for example Refs.~\citenum{mieda:16} for more detail about SHWFS elongation and imperfection in centroiding.

To simulate the SHWFS elongation created by the Na layer, we apply a set of defocus commands to DM0 and change the LGS intensity according to the empirically obtained Na profile while the LGS-WFS camera shutter is open (or take individual WFS image and combine all). An example of the empirical Na profiles taken by the University of British Colombia group is shown in Figure \ref{fig:naprofile}, and its reproduced SHWFS elongation at $t=0$ second is shown in Figure \ref{fig:elon_wfs}.

%======================================
%=== PYRAMID TRUTH WAVEFRONT SENSOR ===
%======================================
\subsection{Pyramid Truth Wavefront Sensor}
\label{sec:pwfs}
The one major update to HeNOS is the implementation of a TWFS made with a PWFS. This is to follow up on the NFIRAOS' new decision to use a PWFS instead of a SHWFS for its TWFS (Ref.~\citenum{veran:15}). The description of general TWFS functions and the HeNOS TWFS design are reported in Refs.~\citenum{mieda:16}, and our optical design is shown in Figure \ref{fig:henos} in green. In short, on HeNOS, the grid of NGSs hitting a star selection mirror (SSM) at the pupil is sent to a pinhole where only one NGS goes through. The single NGS beam is then modulated by a fast steering mirror (FSM) at the pupil, and the focused beam makes a circle around the vertex of the pyramid. The light is distributed in four directions, and a relay lens behind the pyramid forms four separate pupil images on a detector. The incoming wavefront can be measured by comparing their intensity patterns. 

Our SSM and FSM use a Zaber Motorized Gimbal Mount and a Newport FSM-300, respectively. While the Zaber controller is run by USB, the FSM-300 controller requires analog inputs, and thus we installed the same ADLINK data acquisition card (DAQe-2502) as the laser diodes in the HeNOS computer. The interface between the DAQ card and the FSM controller is an off-the-shelf ADLINK terminal board. The PWFS camera is a Point Grey Flea (648$\times$488 array with 5.6 $\mu$m pixel). To evaluate the PSF at the tip of the pyramid, we have inserted one more beamsplitter between the focusing lens and the pyramid and added one more camera (pyramid science camera, PSC, orange path in Figure \ref{fig:henos}). The PSC is a Point Grey Grasshopper (2448$\times$2048, 3.45 $\mu$m pixel). Both the PWFS camera and the PSC are also connected to the ADLINK terminal board so that the PWFS camera, PSC, and FSM are all synchronized.

Our pyramid component is a double pyramid borrowed from the Arcetri group. A double pyramid consists of two pyramid-shaped prisms glued back to back. The two prisms are made with different materials (i.e., different index of refraction) to compensate for chromatic aberrations. Alternative pyramid components, such as double roof prisms, can be used (see e.g. Refs.~\citenum{mieda:16} for performance comparisons between different pyramid components).
\begin{deluxetable}{cccc}
\tabletypesize{\footnotesize}
\tablecolumns{4}
\tablewidth{0pc}
\tablecaption{As-built HeNOS Parameters}
\setlength{\tabcolsep}{0.02in}
\tablehead{
\colhead{Parameter} & 
\colhead{Unit} &
\colhead{Design} & 
\colhead{measurement}
}
\startdata
LGS asterism & [arcsec] & 4.5 & 4.5 \\
Telescope size & [m] & 8 & 8.13 \\
Actuator distance & [m] & 0.89 & 0.914 \\
Subaperture size & [m] & 0.27 & 0.267 \\
Science FOV & [arcsec] & 10.9 & 11.04 \\
DM altitude & [km] & [0, 11.2] & [0, 12] \\
$r_0$(500 nm) & [m] & 0.609 & 0.584 \\
Phase screen altitude & [km] & [0, 4.2, 14] & [0.6, 5.2, 16.3] \\
Phase screen strength & [\%] & [72.3, 19.8, 7.9] & [74.3, 17.4, 8.2] \\
LGS altitude & [km] & 90 & 98.5 \\
\enddata \label{tab:summary}
\end{deluxetable}

%=====================
%=== Bench Summary ===
%=====================
\subsection{Bench Summary}
\begin{deluxetable}{ccccc}
\tabletypesize{\footnotesize}
\tablecolumns{5}
\tablewidth{0pc}
\tablecaption{Desired versus obtained objective}
\setlength{\tabcolsep}{0.02in}
\tablehead{
\colhead{Objective} &
\colhead{Metric} & 
\colhead{Desired} &
\colhead{Obtained} &
\colhead{Error}
}
\startdata
1 & $r_0/d_{\rm H}$ & 1.0 & 0.92 & 8 \% \\
3 & FOV/$\theta_0$ & 12.8 & 9.8 & 23 \% \\
2 & FOV/$\theta_2$ & 3.5 & 2.9 & 17 \% \\
2 and 3 & $\theta_2/\theta_0$ & 3.7 & 3.4 & 8\% 
\enddata \label{tab:obj}
\end{deluxetable}
All calibration results described in this section are summarized in Table \ref{tab:summary}. For simplicity, the values here are not including the stretch factor, $f_s$, described in \S \ref{sec:bench}. Table \ref{tab:obj} summarizes how the as-built bench matches the design objective from \S \ref{sec:bench}. In general, the parameters of the HeNOS bench are in reasonable agreement with the desired parameters. In any case, the most important is to know the as-built parameters so that the models can be parameterized correctly. Knowing the HeNOS parameter well, we are ready to demonstrate and test the techniques that will be used on NFIRAOS and other instruments/AO systems. Please refer additional HeNOS tests in Refs.~\citenum{turri:17} and \citenum{rosensteiner:16} for a SLODAR experiment and NCPA calibration using the focal plane sharpening method.

%========================
%=== Closing the Loop ===
%========================
\section{Closing the Loop}
\label{sec:loop}
Right now, we use a flat mirror at the DM1's position to simplify the calibration procedures and bench development, and thus the HeNOS bench only supports single-conjugate adaptive optics (SCAO). With the SHWFS and the newly implemented PWFS, HeNOS offers three different modes: LGSAO with a SHWFS, NGSAO with a PWFS, and LGSAO with an elongated SHWFS corrected by a NGS-PTWFS. For all modes, we simulate each frame step by step instead of running the bench in real time so that we can isolate and understand the effect of individual step/component. When we simulate each step, for example 1) turn on LGS laser and 2) take SHWFS image in Figure \ref{fig:LGSAO_SHWFS}, we include extra waiting time in each step to make sure the laser is fully lased, and the SHWFS image completely transferred to computer. Because of these waiting time, simulating frames takes long time. Study of hardware latency, such as DM and WFS latency, and optimization of real time controller will be a separate experiment in the future.

%========================
%=== LGSAO with SHWFS ===
%========================
\subsection{LGSAO with SHWFS}
\label{sec:loopshwfs}
\begin{wrapfigure}{r}{0.5\textwidth}
\includegraphics[width=0.5\textwidth]{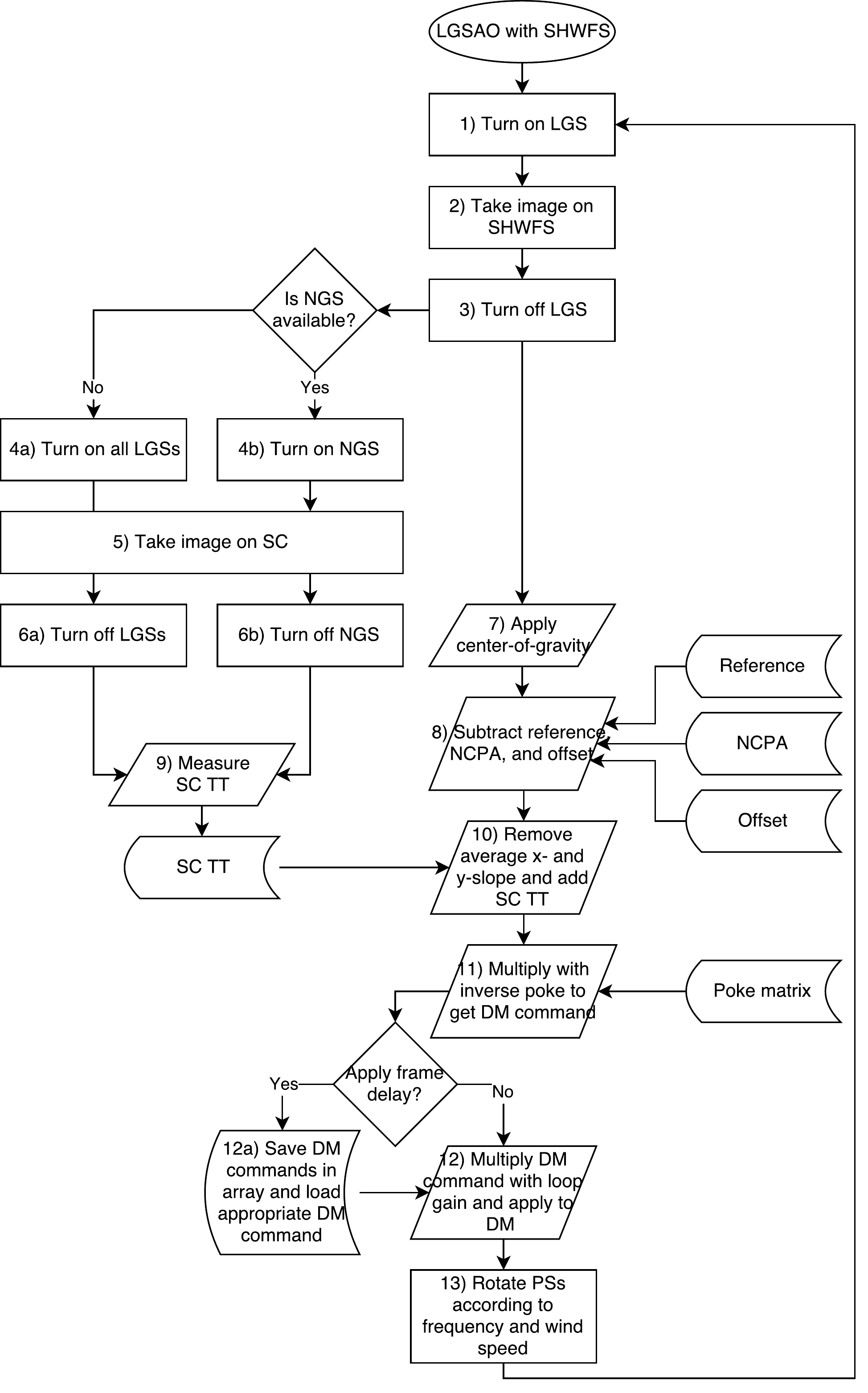}
\caption{Flowchart describing how the SCAO loop is closed using the LGS and the SHWFS on HeNOS. The numbers indicate the order of process called in the loop.}
\label{fig:LGSAO_SHWFS}
\end{wrapfigure}

The steps to simulate the classical SCAO with LGS and SHWFS are shown in Figure \ref{fig:LGSAO_SHWFS}. %For each frame, the bench 1) turns on one LGS laser, 2) takes an image on the SHWFS camera, 3) turns off the LGS laser, 4) turns on all LGSs or NGS (depending on the NGS source availability), 5) takes an image on the SC, 6) turns off the LGS/NGS, 7) applies the center-of-gravity centroiding method to the SHWFS image, 8) applies the position-to-slope conversion factor and subtract the reference, NCPA (if applicable), and offset (if applicable) slopes, 9) measures the science tip and tilt (TT) using the SC image taken at step 5, 10) removes the average x-slope (WFS tip) and the average of y-slope (WFS tilt) from the WFS slopes and add the science TT to the WFS slopes, 11) multiplies the slopes with the inverse of the poke matrix to get the next DM command, 12a) when a frame delay is applied, saves the DM command in an array and load an appropriate (current or few frames back) DM command, 12) multiplies the DM command with a loop gain and applies to DM, and 13) rotates the PSs according to the loop frequency and wind speed. 
The numbers in the flowchart indicate the order of process called in the loop. Currently, it takes about 3 seconds to take one frame on the bench, which leads to several hours of run-time for any experiment. We noticed that over this time, the Strehl ratio of the long 
exposure image drops while the instantaneous Strehl stays stable. This is indicative of a TT error at the WFS. The cause of this TT error is probably produced by the flexure in the paths. In the WFS and science paths on HeNOS after DM0, optical components, including WFS camera, are mounted in between four rods (called cage system) that are more keen to thermal expansion. To correct this, we neglect the TT from the LGSWFS %even in the case of SCAO 
and use the TT from the SC (step 9 and 10 in Figure \ref{fig:LGSAO_SHWFS}). This method prevents TT errors at the SC: the long exposure SR no longer drops and remains comparabole to the average short exposure SR.

This mode is the most developed and most used mode on the HeNOS bench. Using this mode, we collaborate with TMT Observatory Corporation and Laboratoire d'Astrophhsique de Marseille to demonstrate PSF reconstruction techniques by comparing the empirical data with analytic models (Refs.~\citenum{gilles:17} and \citenum{martin:18}). 

%=======================
%=== NGSAO with PWFS ===
%=======================
\subsection{NGSAO with PWFS}
\label{sec:looppwfs}
\begin{wrapfigure}{r}{0.5\textwidth}
\includegraphics[width=0.5\textwidth]{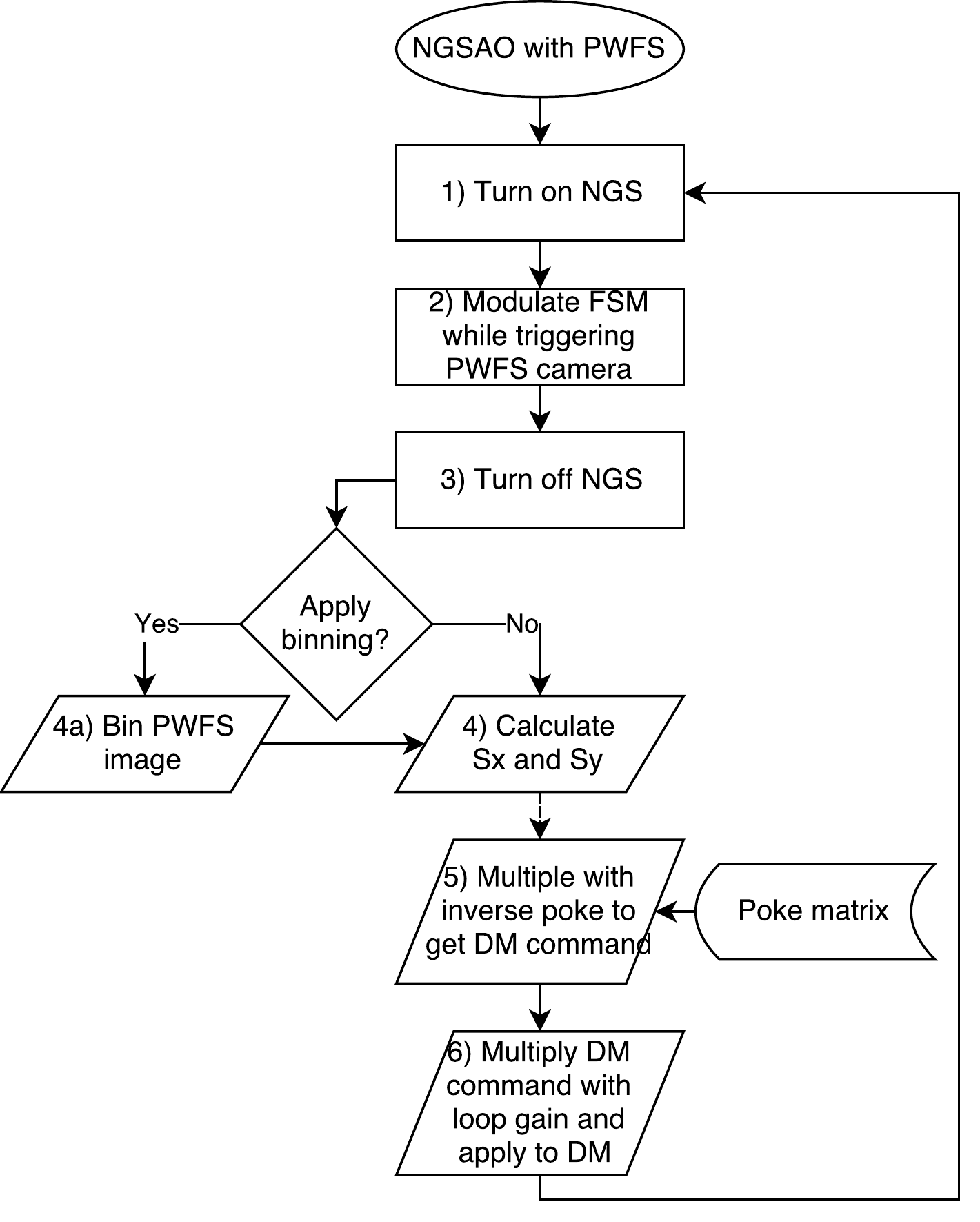}
\caption{Flowchart describing how the HeNOS bench closes the loop with NGS and PWFS. The numbers indicate the order of process called in the loop.}
\label{fig:NGSAO_PWFS}
\end{wrapfigure}

With the implementation of the PWFS, the bench now can be closed with the NGS and the PWFS. The steps of this mode are shown in Figure \ref{fig:NGSAO_PWFS}.% and described as follows: 1) turn on the NGS, 2) modulate the FSM while triggering the camera, 3) turn off the NGS, 4a) apply binning to the PWFS image if appreciable, 4) calculate the x- and y- signal (Sx and Sy), 5) multiply with the inverse of the poke matrix to get a DM command, and 6) multiply the DM command with a loop gain and apply to DM0.Because this mode is not our main interest, the current development is quite basic.

\begin{figure}[b]
\centering
\includegraphics[width=0.99\textwidth]{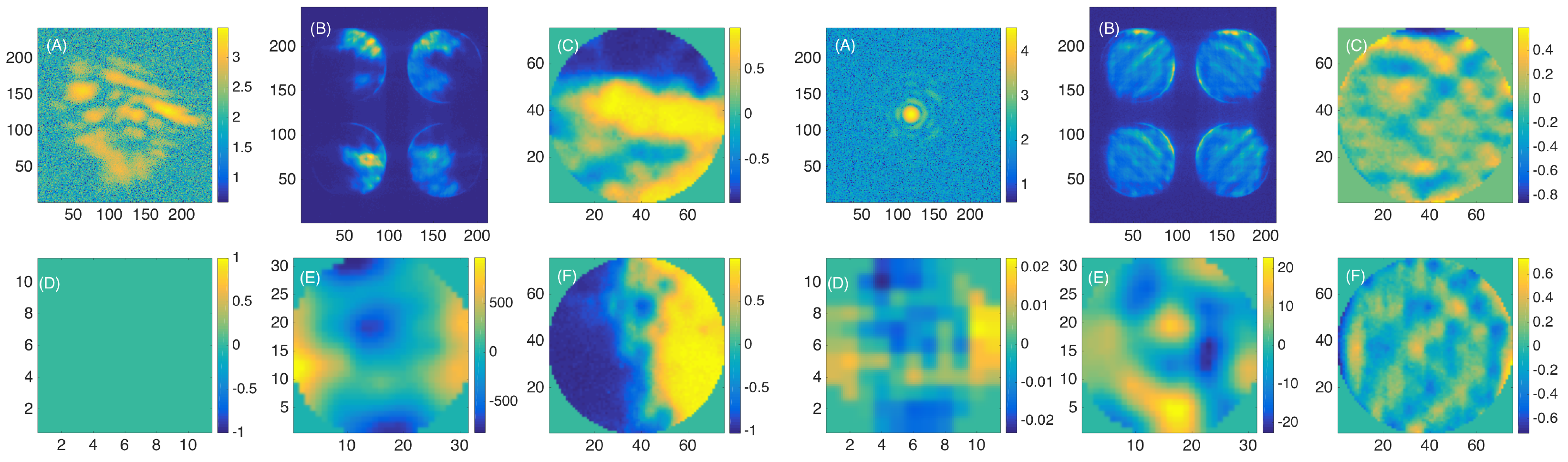}
\caption{(A) PSF on PSC, (B) raw pupil image, (C) x- and (F) y-slope signal, (D) DM0 shape, and (E) reconstructed wavefront. Starting from zero command (6 panels on the left), the loop is closed after 50 iterations with loop gain = 0.2 (6 panels on the right), and diffraction rings are seen in the PSF (panel A).}
\label{fig:pwfs50}
\end{figure}

As an example, (A) PSF, (B) raw PWFS image, (C) x- and (F) y-slope signal, (D) DM0 shape, and (E) reconstructed wavefront before (6 left panels on Figure \ref{fig:pwfs50}, zero commands on DM0) and after (6 right panels) the loop is closed are shown. In this example, there are no PSs in the path. The initial PSF is big and fuzzy, and after closing the loop, the diffraction rings are visible. 

Because the main purpose of the PWFS is a TWFS, this NGSAO with PWFS mode is currently quite basic. Once the new NGS source with a larger pinhole is added (\S \ref{sec:ngs}), we will experiment wavefront sensing on an extended object with PWFS. Extended wavefront sensing in general is useful to increase the sky coverage, but also its application includes the use of LGS on a PWFS.
%}
As PWFS becomes a more popular choice for many future upgrades (e.g., GPI, Subaru, and Keck), the PWFS mode on the HeNOS bench is an interesting capability for future experiments.

%================================================
%=== LGSAO with elongated SHWFS and NGS PTWFS ===
%================================================
\subsection{LGSAO with elongated SHWFS and NGS PTWFS}
\label{sec:lgstwfs}
\begin{wrapfigure}{r}{0.5\textwidth}
\centering
\includegraphics[width=0.5\textwidth]{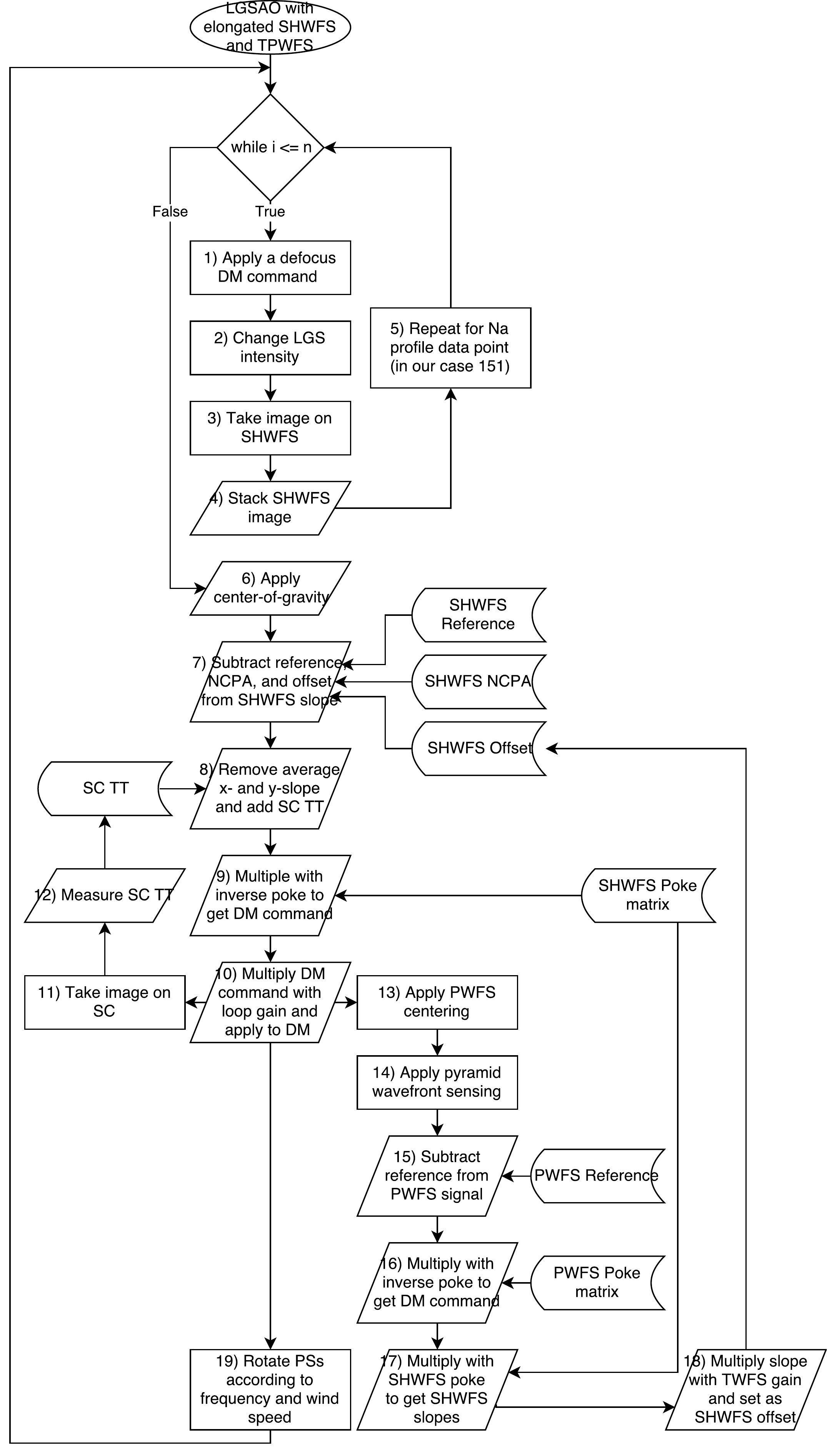}
\caption{Flowchart describing how the elongated SHWFS is integrated into the SCAO loop and how measurements from the PTWFS are fed to the loop as SHWFS offsets. The numbers indicate the order of process called in the loop.}
\label{fig:LGSAO_PTWFS}
\end{wrapfigure}

The last mode available on the HeNOS bench is the loop with the LGS-SHWFS using the elongated SHWFS spots and the NGS-TWFS correction. The flowchart in Figure \ref{fig:LGSAO_PTWFS} shows the steps. %The loop is closed as follows: 1) apply a given amount of defocus to DM0, 2) turn on a LGS with a corresponding intensity, 3) take an image on the SHWFS, 4) stack the image, 5) repeat the step 1 through 4 for the number of Na profile data points (in our case 151 points), 6) apply the center-of-gravity centroid method to the final stack image, 7) apply the position-to-slope conversion factor and subtract the reference, NCPA (if applicable), and offset (if applicable) slopes, 8) removes the average x- and y-slope and add the SC TT, 9) multiply the slope with the inverse of the SHWFS poke matrix to get a DM command, 10) apply the new DM command multiplied by a loop gain to DM0, 11) take a science image, 12) measure a TT from the science image and save the value to be used in the step 8, 13) apply a PWFS centering, 14) apply the pyramid wavefront sensing described in the step 1 through 5 in S \ref{sec:pwfs}, 15) subtract the reference PWFS signal, 16) multiply with the inverse of the PWFS poke matrix to get a DM command, 17) multiply the new DM command with the SHWFS poke to get the SHWFS slope, 18) multiply this SHWFS slope by a TWFS gain and add to the current offset for the elongated SHWFS, and 19) rotate the PSs according to the loop frequency and wind speed.

Because of the alignment imperfection in the PWFS path, as the amount of defocus applied on DM0 changes, the position of the beam at the tip of the pyramid drifts. To fix this problem, we update the zero-position of FSM each time (PWFS centering at the step 13). The new zero-position is measured by comparing the vertical and horizontal pupil intensities. As we apply the evolving Na profile (Figure \ref{fig:naprofile}), the mean height of Na layer (thus focus term) need to be measured. In the TMT+NFIRAOS case, the TWFS does not measure the defocus term and only measures higher order radial modes, but on the HeNOS bench, we measure the defocus term using the TWFS as well because we do not have a separate focus wavefront sensor.

\begin{figure}[t]
\centering
\includegraphics[width=0.99\textwidth]{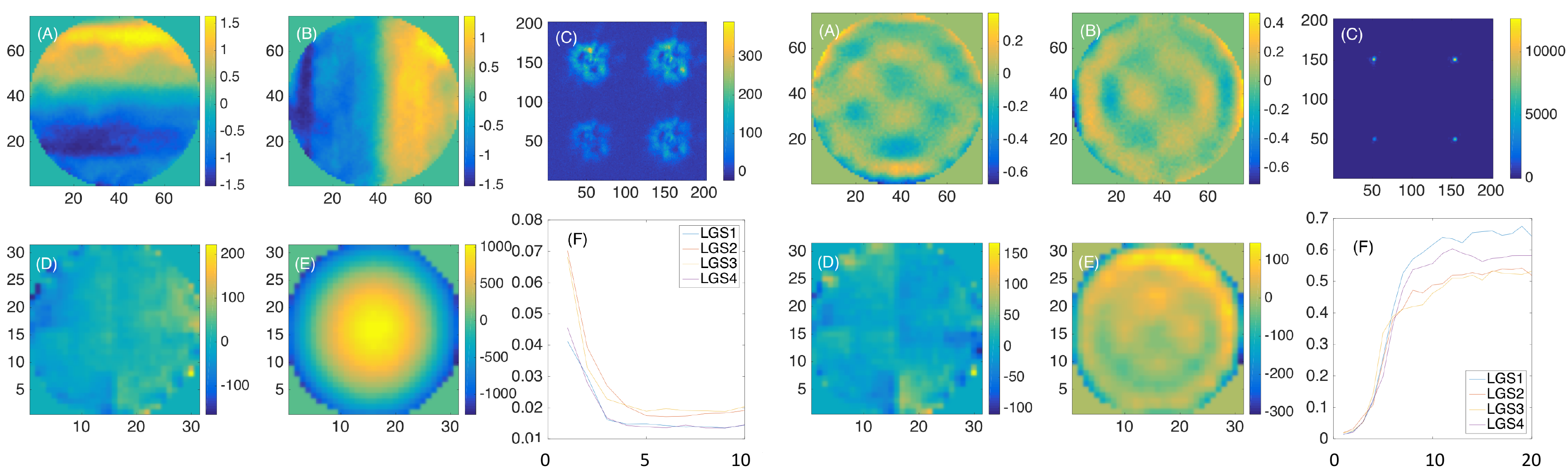}
\caption{(A) PWFS slope signal in x, (B) PWFS slope signal in y, (C) four LGSs on the science camera, (D) reconstructed phase using elongated SHWFS, (E) reconstructed phase using normal SHWFS, (F) evolution of four LGS Strehl ratios while closing loop. 6 panels on the left: After closing loop on the elongated SHWFS spots with the Na profile shown in the right panel of Figure \ref{fig:naprofile} without TWFS feedback. 6 panels on the right: After closing loop with TWFS feedback}.
\label{fig:withtwfs}
\end{figure}

Figure \ref{fig:withtwfs} shows %the result
one example of closing loop with the elongated SHWFS. Starting with the best DM flat from the previous day, the loop is closed using the elongated SHWFS spots. For this experiment, we apply the same Na profile
shown in Figure \ref{fig:naprofile} on the right for all iterations (i.e., Na profile does not evolve). We use a higher loop gain of 0.5 for this test to speed up the simulation. Before the TWFS feedback is applied (6 left panels on Figure \ref{fig:withtwfs}), DM changes its shape to compensate the aberrations created in the elongated SHWFS spots by the Na profile, as described in \S \ref{sec:elon}, and thus the reconstructed phase by the elongated SHWFS (panel D) sees flat phase. However, the true wavefront on the science targets are not affected by the Na profile (the actual phase shown on panel E includes a big defocus and radial aberrations), and it actually adds "wrong" aberration on to the science targets. As a result, four LGS Strehl ratios (panel F) become low, and the final PSFs on panel C are big and fuzzy. %because the loop is closed to the Na profile, the reconstructed phase by the elongated SHWFS (panel D) sees flat phase while the actual phase (panel E) includes a big defocus and radial aberrations. The defocus is also seen on the PWFS (panel A and B). Panel F shows four LGS Strehl ratios become low, and well before the 10th iteration, it plateaus. The final PSFs on panel C are big and fuzzy. 

After the loop is closed on the elongated SHWFS with Na profile, we then close the loop again but now with TWFS feedback (right 6 pannels on Figure \ref{fig:withtwfs}). The TWFS feedback is applied as a reference slope to the elongated SHWFS. After 20 iterations, the four LGS Strehl ratios (panel F) are improved ($>$ 50\%), and all three WFSs, PWFS (panels A and B), elongated SHWFS (panel D), and normal SHWFS (panel E) see small aberrations.

%===============================
%=== SUMMARY AND FUTURE PLAN ===
%===============================
\section{Summary and Future Plan}
\label{sec:sum}
We have shown the derivations of the HeNOS parameters (\S \ref{sec:bench}) and reported the implementation of these parameters on an optical bench (\S \ref{sec:benchcomp}). The most recent upgrade is the addition of the TWFS made with a PWFS (\S \ref{sec:pwfs}). The truth wavefront sensing includes the simulation of the SHWFS spot elongation due to the Na layer (\S \ref{sec:elon}), and PWFS feedback as a reference slope to the SHWFS. With the new implementation of the PWFS, the HeNOS bench currently offers three different AO modes: LGSAO with a SHWFS (\S \ref{sec:loopshwfs}), NGSAO with a PWFS (\S \ref{sec:looppwfs}), and LGSAO with an elongated SHWFS corrected by a NGS-PTWFS (\S \ref{sec:lgstwfs}). %An 
Some examples of actual experiment performed on HeNOS can be found in Ref.~\citenum{gilles:17} and Ref.~\citenum{martin:18}, where we use HeNOS to demonstrate NGAO PSF reconstruction algorithm for TMT-NFIRAOS and a generalized off-axis PSF reconstruction model for extremely large telescopes, respectively.

There are several hardware and software additions to be completed soon. For hardware additions, we have started to work on a new NGS source simulator design using a pinhole mask (\S \ref{sec:ngs}). We will make more measurements with the prototype (Figure \ref{fig:pup} bottom panel) and finalize the design. Once the new NGS light source is built, we can move the science camera back to the NGS focus position. Currently, we use a flat mirror instead of a DM at the DM1's position to simplify the bench for calibration purposes. Most calibrations are done, and we expect to implement DM1 soon.

For the software upgrades, we are working on including a matched filter centroiding method for the LGS-WFS. It is the method NFIRAOS will use. Currently, we close the loop only with one LGS, thus it is a SCAO loop. We are collaborating with Laboratoire d'Astrophysique de Marseille on implementing a laser tomography (LT) AO functionality to HeNOS, which can be achieved without having the second DM. LTAO mode can potentially be operated on NFIRAOS, especially when one of the two DMs fails. Once we add the second DM, we can also include multi-conjugate AO. With the LTAO and MCAO capabilities, PSF reconstruction for LTAO/MCAO will also be tested.

Once all upgrades are included, the bench will not only provide an experimental anchor to the model currently used to predict the NFIRAOS performance, but will also demonstrate the followings:
\begin{itemize}
    \item an MCAO configuration where the matched filter for LGS-SHWFS centroiding is updated according to changes in the sodium (Na) layer profile using one TWFS.
    \item robustness against the spatial non-uniformity of the Na layer, where each WFS sees a different Na profile.
    \item an MCAO configuration where field-dependent non-common path aberrations (NCPAs) are calibrated and compensated with turbulence, LGS elongation, and Na profile evolution via WFS slope offsets, on top of off-line (no spot elongation and no turbulence) calibration.
    \item tomographic reconstruction using a combination of high-order LGS-WFS and low-order NGS-WFS, particularly with NGSs that are faint and/or only partially sharpened by the MCAO system.
\item turbulence profile estimation using techniques similar to SLODAR (SLOpe Detection And Ranging; Refs.~\citenum{wilson:02, butterley:06}) applied to the NFIRAOS LGS-WFS measurements.
    \item wide-field point spread function (PSF) reconstruction in an MCAO system.
\end{itemize}

Calibration methods, algorithms, and AO techniques developed here are all valuable % knowledge
not only for NFIRAOS development but also for future upgrades and developments of other instruments. We will continue developing the bench to include more functionalities to simulate NFIRAOS while collaborate with others to experiment their new techniques. 

\acknowledgments
The development of the HeNOS bench has been a collective effort of many members, including short term postdoctoral researchers and students, and we would like to acknowledge Masen Lamb at Dunlap Institute at University of Toronto, Siqi Liu at University of Toronto, Maaike van Kooten at Leiden University, Carlos Correia at Laboratoire d'Astrophysique de Marseille, and Eric A. McVeigh for their involvement. We also would like to thank Tim Hardy and Jonathan Stocks at NRC-Herzberg for their generous technical help in hardware and software. 

\bibliography{reference}
\bibliographystyle{spiejour}

\end{document}